\documentclass[prb,twocolumn,floatfix,showpacs,superscriptaddress]{revtex4-1}
\usepackage{graphicx,epsfig}

\begin{document}

\title{Screened Coulomb Hybrid Density Functional Investigation of Oxygen 
Point Defects on ZnO Nanowires}
\author{Veysel \c{C}elik}
\affiliation{Department of Computer Education and Educational 
Technologies, Siirt University, Siirt 56100, Turkey}
\author{Ersen Mete}\email{emete@balikesir.edu.tr}
\affiliation{Deparment of Physics, Bal{\i}kesir University,
Bal{\i}kesir 10145, Turkey}

\date{\today}

\begin{abstract}
In this study, oxygen vacancies and adatoms have been considered on
the surface of both hexagonal and triangular ZnO nanowires. Their effect on 
the electronic structure and optical spectra of the nanowires have been
investigated using the exact exchange hybrid density functional theory 
calculations. A surface oxygen vacancy gives rise to appearance of 
a band gap state at almost 0.7 eV above the valence band of the both types of 
the nanowires while an oxygen adatom show bulk-like electronic properties. 
A shape dependence is also indicated by the calculated physical quantities 
of oxygen related point defects on ZnO nanowires.
\end{abstract}


\maketitle

\section{Introduction}
Today, the need for clean energy sources is increasing day by day. In this 
sense, solar cells are an important way of obtaining energy. Excitonic solar 
cells\cite{Gregg} are promising devices for its low-cost and photocatalytic 
properties. In terms of efficiency and stability, dye-sensitized solar cells 
(DSSCs)\citep{Gratzel} are promising systems among the excitonic solar 
cells.\citep{Nazeeruddin,Wang} Generally, in DSSCs, thick nanoporous films 
of TiO$_2$ or ZnO are used as the anode.\citep{Keis,Gratzel} Because of 
uneven structure and thickness of these films, trap states occur which 
shorten diffusion length of electrons in the oxide.\citep{Law} Moreover, 
presence of such a trapping mechanism increases the probability of 
electron-hole recombination rate causing a significant reduction in the 
solar-to-electricity conversion efficiency in a real world application. Law 
\textit{et al.}~\cite{Law} introduced a version of the DSSCs in which the 
traditional nanoporous film is replaced by a dense array of oriented, 
crystalline ZnO nanowires. The study shows that replacing the nanoparticulate 
film with an array of well oriented single-crystalline nanowires increases the 
electron diffusion length in the anode. For photocatalytic applications, the 
nanowire (NW) form has many advantages such as providing a natural pathway for 
the photo-excited electrons to flow through and allowing a higher 
surface-to-volume ratio for accommodating larger number of light harvesting 
chromophore adsorbates.\citep{Ong,Zhang}

\begin{figure}[b]
\epsfig{file=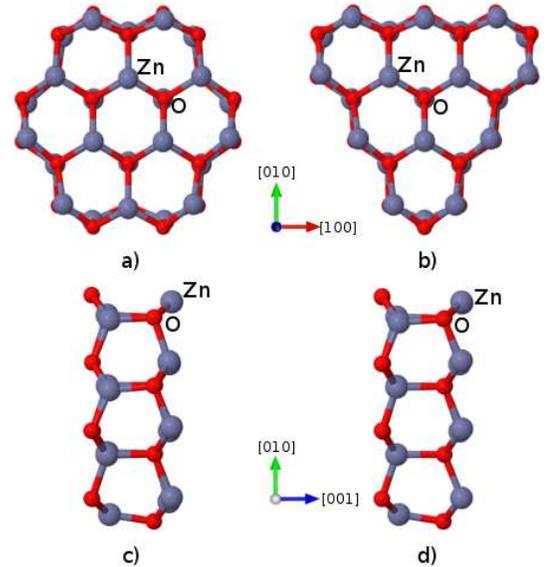,width=7cm}
\caption{Optimized structures in the ball-and-stick representation of the 
hexagonal and the triangular [0001] ZnO nanowire models. Cross sectional views 
(a) and (b) through the [001] direction and side views (c) and (d) along the 
[001] direction are presented for the hexagonal and the triangular cells, 
respectively.\label{fig1}}
\end{figure}

Studies report an unintentional n-type conductivity in ZnO depending on 
growth processes. One of the reasons can be the existence of 
shallow-donor impurities which might be involved in the 
samples.\cite{Janotti,Lany2005,Liu} The point defects such as oxygen vacancies 
($V_{\textrm{\scriptsize O}}$) were believed to be not contributing to the 
n-type conductivity.\cite{Janotti} That is due to the fact that 
$V_{\textrm{\scriptsize O}}$ is a deep rather than a shallow donor. On the 
other hand, Lany \textit{et al.}\cite{Lany2005} used GW calculations 
to predict that under illumination a shallow metastable state appear associated 
with a charge oxygen state which could contribute to n-type conductivity in 
bulk ZnO. Similarly, Liu \textit{et al.}, in a recent experimental study, 
conclude that the oxygen vacancy $V_{\textrm{\scriptsize O}}$ at a charge 
state of +2 is the main reason of n-type conductivity in ZnO 
material.\cite{Liu} The $V_{\textrm{\scriptsize O}}$ is a native defect 
and its formation energy is lower than that of a zinc 
interstitial.\cite{Oba} The formation and the effect of 
$V_{\textrm{\scriptsize O}}$ on the properties of ZnO have been experimentally 
studied by many 
groups.\cite{Casteleiro,Look,Lany2005,Lany2007,Lany2008,Lany2010,
Janotti2007,Paudel,Pemmaraju,Vines1,Repp,Wang2012,Janotti2005,Leiter,Hofmann,
Li2014} Oxygen vacancies are also reported for ZnO 
nanowires (NWs).\cite{Muchuweni,Su,Wang2008,Sheetz,Deng} 
Therefore, understanding the role of the oxygen related defects on the 
electronic and optical properties of ZnO nanowires is important. 
Moreover, the effect of defect states on the band gap associated properties of 
ZnO NWs must be clarified in relation to DSSCs. Therefore, one of the focuses
of this study is the electronic structure of ZnO nanowires with oxygen defects. 

Under normal conditions, ZnO crystallizes in a hexagonal wurtzite phase. Lattice 
parameters of ZnO are reported as  $a$ = 3.2475 {\AA} and $c$ = 5.2066 
{\AA}.\cite{Gerward} For bulk ZnO, the lattice parameters calculated with the 
DFT are in good agreement with the experimental data.\cite{Haffad,Ozgur} 
However, due to local density approximations to the exchange-correlation (XC) 
effects included in the standard DFT, the GGA type functionals lead to severely 
underestimated electronic band gaps. ZnO has a direct band gap of about 3.4 
eV\cite{Kumar,Reynolds} whereas GGA-DFT calculations reported a value of
about 0.74 eV.\cite{Haffad,Oba} In addition, the position and nature of 
probable energy levels in the band gap of ZnO associated with various types of 
defects and impurities can not be correctly predicted by the standard XC 
functionals. Therefore, hybrid DFT methods, which partially incorporate exact 
exchange, can be used to overcome these issues. A recent study shows that 
hybrid Hartree-Fock DFT calculations can be successful in describing the 
defect energetics and the electronic structure of bulk ZnO with 
impurities.\cite{Oba}

Fabricated ZnO nanowires are generally in hexagonal 
structure which usually have oxygen deficiencies.\cite{ZLWang2008,Muchuweni,Su} 
Moreover, with thermal evaporation processes, nanowires in triangular structure 
can also be synthesized.\cite{Wang2006} Defect-free ZnO NWs were studied by 
DFT calculations including supplementary Hubbard $U$ terms which still shows 
underestimation in the predicted band gaps.\cite{Haffad} 
In this study, we used the screened exchange hybrid density functional theory 
(DFT) method to get a proper description of the electronic structures and 
optical spectra of hexagonal and triangular ZnO nanowires with surface oxygen 
defiencies and adatoms.

\begin{figure}[b]
\epsfig{file=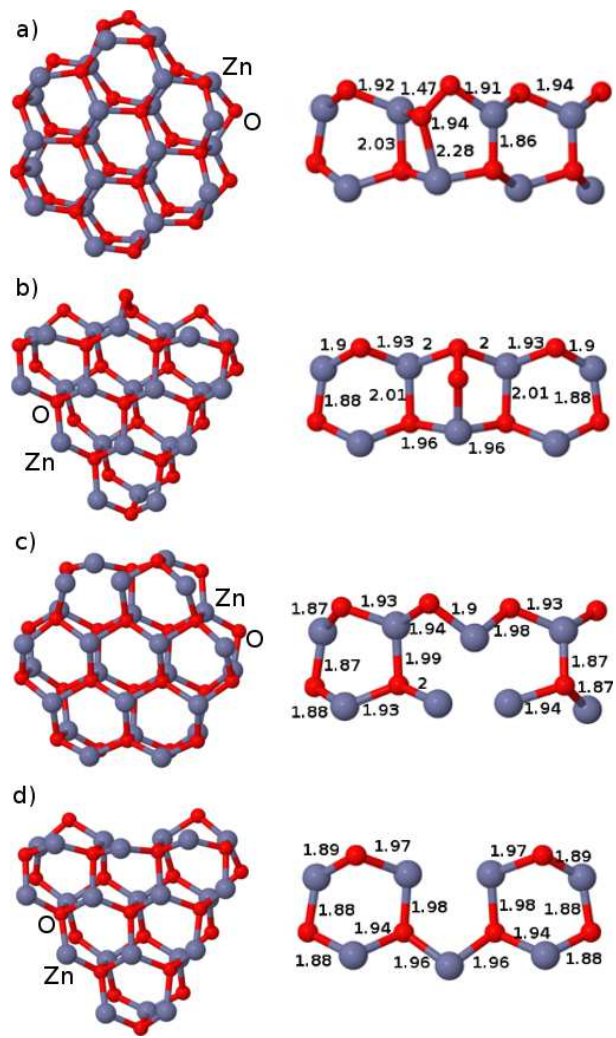,width=8cm}
\caption{The unit cell structures of (a) the hexagonal and (b) the triangular 
ZnO nanowires with an O adatom optimized using the HSE functional. The O 
vacancy structures are depicted for (c) the hexagonal and (d) the triangular 
NWs. Bond lengths are given in units of {\AA}.\label{fig2}}
\end{figure}

\begin{figure*}[htb]
\epsfig{file=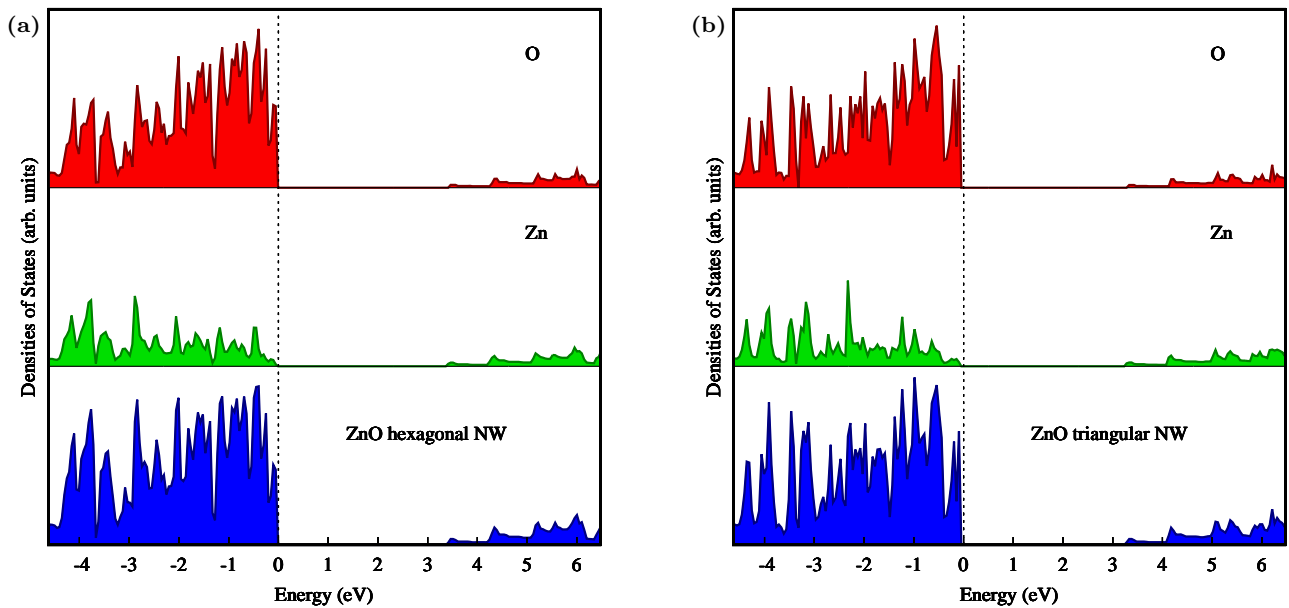,width=17cm}
\caption{The HSE-calculated partial and total densities of states (DOS) of 
(a) hexagonal and (b) triangular ZnO nanowires. The bottom panels show
the total DOS structures. The dotted vertical lines denote the Fermi energies 
and set at slightly above the highest occupied states.\label{fig3}}
\end{figure*}

\section{Computational Method}

Periodic total energy density functional theory (DFT) calculations have been 
performed based on the projector-augmented wave (PAW)\cite{Blochl,Kresse2}
method as implemented in the Vienna ab-initio simulation package 
(VASP).\cite{Kresse1,Kresse3} A kinetic energy cutoff of 400 eV was used to 
expand single particle states in plane-waves. The exchange-correlation effects 
have been taken into account by employing the range separated hybrid HSE 
functional.\cite{HSE03,Paier2006}

The lack of a proper self-interaction correction (SIC) leads to 
the well-known band gap underestimation by the standard DFT 
exchange-correlation functionals, such as PBE.\cite{PBE1996}
On the other hand, Hartree-Fock (HF) formalism has well-defined Coulomb
direct and exchange terms canceling each other for the zero momentum 
components avoiding self-interaction of charges. In order to benefit 
from this, modern hybrid DFT functionals partially admix the nonlocal exact 
exchange energy with the semilocal PBE exchange energy. 

The hybrid HSE functional treats the exchange energy as composed of long-range 
(LR) and short-range (SR) parts with a range separation parameter $\omega$ and 
mixes the exact exchange with the PBE exchange at the short-range by a mixing 
factor of $a=0.25$ such that,\cite{HSE03,Paier2006}
\[
E_{\textbf{\tiny X}}^{\textrm{\scriptsize HSE}}=
a E_{\textbf{\tiny X}}^{\textrm{\scriptsize HF,SR}}(\omega)+
(1-a)E_{\textbf{\tiny X}}^{\textrm{\scriptsize PBE,SR}}(\omega)+
E_{\textbf{\tiny X}}^{\textrm{\scriptsize PBE,LR}}(\omega)\,.
\]
The correlation term of the XC energy is taken from standard PBE correlation 
energy.\cite{PBE1996} Previous theoretical studies show that HSE functional is 
useful to get electronic band gap related features of periodic and finite 
physical systems reasonably accurate to be comparable with 
experiments.\cite{Henderson2011,Celik2012,Vines2} For this reason, we used the 
range-minimized HSE12s~\cite{Moussa2012} variant which lowers the computational 
cost without losing the accuracy of the original HSE06 functional.

Initial stoichiometric triangular and hexagonal [0001] NW models were carved 
from ZnO bulk structure. The computational cells for the hexagonal and 
triangular NW models contain a total of 48 and 44 atoms, respectively. We 
assumed periodic boundary conditions and chose the $x$-direction to be the 
major axis of the NW. Then, the atomic coordinates were optimized 
self-consistently until each cartesian component of the Hellman-Feynman force 
acting on each atom in the cell to be less than 0.01 eV/{\AA}. Symmetry was not 
imposed throughout any of the computations. Moreover, none of the atoms were 
frozen to their bulk positions to allow a full relaxation. In order to 
avoid any spurious interaction between the periodic images of the NW, we 
introduced a vacuum separation of at least 10 {\AA} in both $y$ and 
$z$-directions in the computational unit cell. Brillouin zone integrations were 
carried out over a 14$\times$1$\times$1 $k$-point mesh being compatible with 
the geometry of the computational cell. We also tested the effect of spin 
polarization and found no significant variation on the calculated results.

\begin{table}[h]%[htb]
\caption{The theoretical and experimental structural parameters of ZnO 
wurtzite. The values are given in {\AA}.}
\begin{ruledtabular}
\begin{tabular}{cccccccc}
\multicolumn{1}{c}{Method}&\multicolumn{1}{c}{$a$}&\multicolumn{1}{c}{
$c$}\\[1mm]\hline
PBE   & 3.26~~~ & 5.22~~~ \\
HSE   & 3.23~~~ & 5.19~~~ \\
Experimental\cite{Gerward} & 3.2475 & 5.2066 \\
\end{tabular}
\end{ruledtabular}
\label{table1}
\end{table}

\begin{figure*}[htb]
\epsfig{file=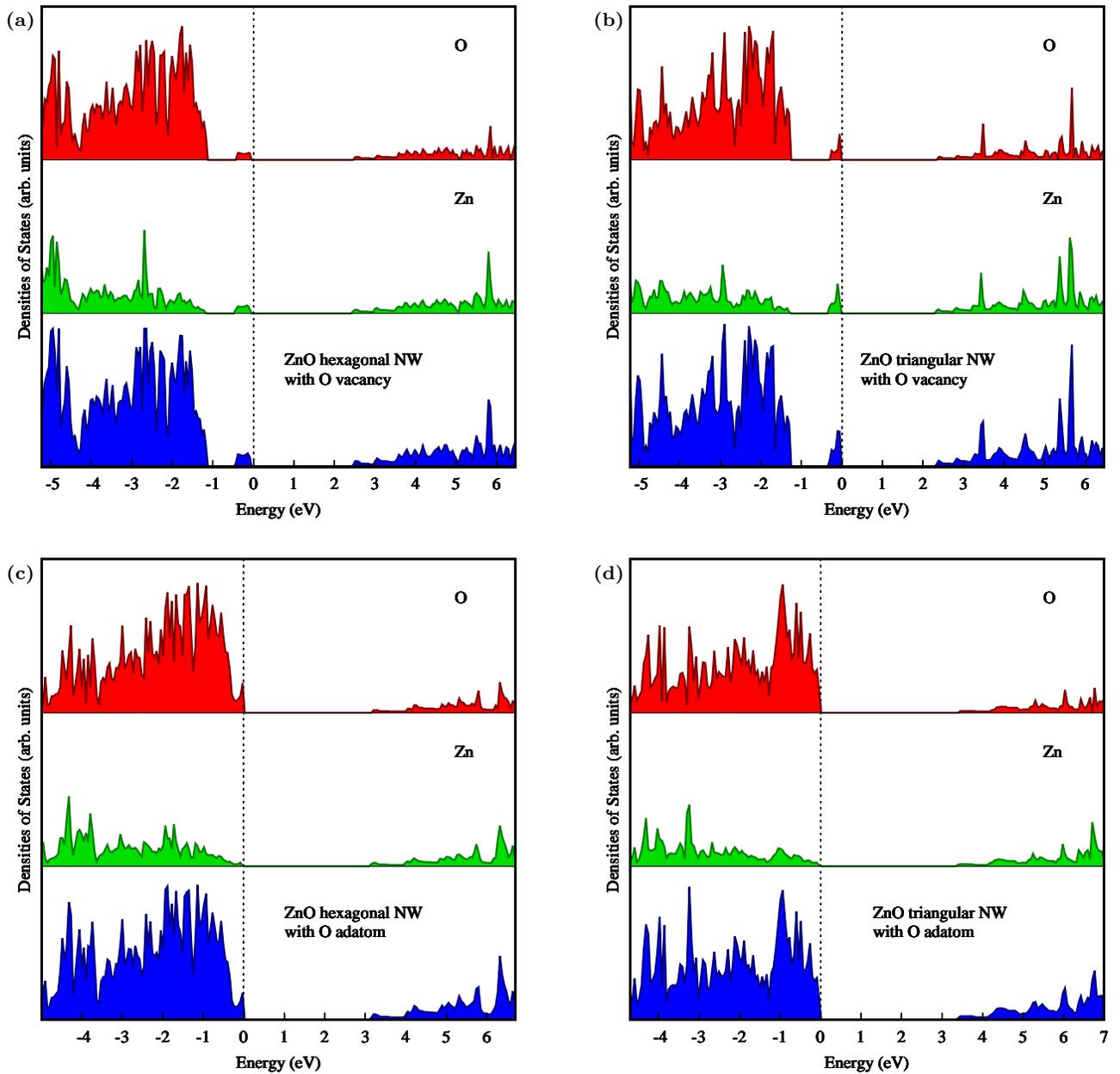,width=17cm}
\caption{The HSE-calculated partial and total densities of states (DOS) of 
the hexagonal ZnO nanowires (a) with a single oxygen vacancy or (c) with a 
single adatom. The results for the triangular nanowire are shown on the right 
(b) for an oxygen vacancy and (d) for an oxygen adatom cases. The bottom panels 
in each figure show the total DOS structures. The dotted vertical lines denote 
the 
Fermi energies.\label{fig4}}
\end{figure*}

\section{Results \& Discussion}

In order to compare theoretically predicted and experimentally estimated 
structural parameters of bulk ZnO in the wurtzite form, the lattice constants 
have been calculated using different XC functionals. The corresponding values 
are obtained as $a$ = 3.26 {\AA}, $c$ = 5.22 {\AA} at the PBE, and as $a$ = 
3.23 {\AA}, $c$ = 5.19 {\AA}  the HSE levels of theory. Although the PBE results 
are fairly good as seen in Table.~\ref{table1}, the HSE functional improves 
theoretical prediction toward a better agreement with the experimental 
values.\cite{Ozgur} 

The enthalpy of formation of bulk ZnO is given by 
\[
\Delta H_{\textrm{\scriptsize f}} = \mu_{\textrm{\scriptsize Zn}} + 
\mu_{\textrm{\scriptsize O}}
\]
where $\mu_{\textrm{\scriptsize Zn}}$ and $\mu_{\textrm{\scriptsize O}}$ are 
the chemical potentials of zinc and oxygen. The computations with the HSE 
functional give $\Delta H_{\textrm{\scriptsize f}}$ for bulk ZnO in the 
wurtzite phase as $-3.52$ eV per Zn that not only agrees well with previous 
theoretical predictions of $-3.4$ eV\cite{Deng,Ganesh} but also shows an 
improvement toward the experimental value of $-3.6$ eV.\cite{Janotti}

\begin{table}[h]%[htb]
\caption{The comparison of computational electronic band gaps of defect-free 
hexagonal and triangular ZnO NWs.}
\begin{ruledtabular}
\begin{tabular}{lcc}
\multicolumn{1}{c}{}&\multicolumn{2}{c}{Band Gap (eV)} \\
\cline{2-3}
\multicolumn{1}{c}{Method}&\multicolumn{1}{c}{Hexagonal}&\multicolumn{1}{c}{
Triangular}\\[1mm]\hline
GGA   &  1.55\cite{Haffad}&1.48\cite{Haffad} \\
HSE   & 3.3~~~~ & 3.2~~~~ \\
\end{tabular}
\end{ruledtabular}
\label{table2}
\end{table}

Firstly, geometry optimization calculations were performed for the 
stoichiometric hexagonal and triangular cross-sectional models. The optimized 
unit cell structures are obtained as shown in Fig.\ref{fig1}, where the average 
Zn-O bond length matches the bulk value of 1.98 {\AA} at the core region. 
However, it reduces to 1.89 {\AA} for the outermost oxygens due to surface 
termination.  The electronic structure calculations using the HSE method 
predicts the band gap values of the stoichiometric hexagonal and triangular ZnO 
NWs as 3.3 eV and 3.2 eV, respectively. The calculated band gap values are 
given in Table.~\ref{table2}. When compared with the previous standart DFT 
results,~\cite{Haffad} the calculated band gap values appears to be largely 
corrected by the HSE functional and agrees very well with the 
experiments.\cite{Leiter,Hofmann,Li2014} The electronic energy gap values 
of the stoichiometric nanowires with diameters about 1 nm as considered 
in this study show, therefore, bulk-like characteristics.

We considered two types of point defects which are probable in an 
experimental environment under oxygen poor or rich conditions. These are 
basically an oxygen vacancy or an adatom cases which can be formed separately 
on the surfaces of both of the NW models. The separation between the periodic 
images of defect sites is 5.17 {\AA} along the nanowire axis, and therefore, 
they can be considered as isolated. The optimized atomic coordinates are shown 
in Fig.~\ref{fig2}. A single point defect causes slight variations on the bonds 
in the local environment. 

In thermodynamic equilibrium, formation energy of a surface oxygen vacancy can 
be calculated in reference to partial oxygen pressure in the environment.
The chemical potentials of Zn and O can vary between their lower and upper 
bounds with respect to each other obeying the enthalpy of ZnO as the stability 
condition of the bulk phase. At O-rich conditions, $\mu_{\textrm{\scriptsize 
O}}^{\textrm{\scriptsize max}}=1/2E(\textrm{O}_2)$ is the energy of O atom in 
an O$_2$ molecule and $\mu_{\textrm{\scriptsize Zn}}^{\textrm{\scriptsize 
min}}=E(\textrm{Zn})+\Delta_{\textrm{\scriptsize f}}(\textrm{ZnO})$ where 
$E(\textrm{Zn})$ is the energy per atom in bulk Zn. O-poor conditions refer to 
the limiting values of $\mu_{\textrm{\scriptsize Zn}}^{\textrm{\scriptsize 
max}}=E(\textrm{Zn})$ and $\mu_{\textrm{\scriptsize O}}^{\textrm{\scriptsize 
min}}=1/2E(\textrm{O}_2)+\Delta_{\textrm{\scriptsize f}}(\textrm{ZnO)}$
Then, the formation of a surface oxygen vacancy on the ZnO NW is
\[
 E_{\textrm{\scriptsize 
f}}(V_{\textrm{\scriptsize O}})=E(\textrm{ZnONW:V}_{\textrm{\scriptsize 
O}})-E(\textrm{ZnONW})+\mu_{\textrm{\scriptsize O}}\
\]
where $E(\textrm{ZnONW:V}_{\textrm{\scriptsize O}})$ and $E(\textrm{ZnONW})$ 
are the total computational cell energy of ZnO NW with and without a charge 
neutral oxygen vacancy, respectively. Under O-rich conditions,
$E_{\textrm{\scriptsize f}}(V_{\textrm{\scriptsize O}})$ gets values as high as 
5.71 eV for the hexagonal and 5.80 eV for the triangular NWs. These 
results agree well with the previous estimations based on 
theory\cite{Ganesh} and experiments\cite{Kim} in the case of bulk ZnO.
At the other limit, O-poor conditions must be considered. Komatsuda \textit{et 
al.} experimentally estimated the oxygen vacancy formation energy in bulk ZnO 
as almost 0.72(6) eV.\cite{Komatsuda} In the case of bulk ZnO theoretical 
predictions show a good agreement with the experiments.\cite{Deng} Our HSE 
calculations give 0.17 eV and 0.08 eV for the formation of a surface oxygen 
vacancy on the hexagonal and triangular ZnO NWs, respectively. Deng \textit{et 
al.} used the PBE functional on a hexagonal ZnO NW with a larger diameter and 
reported that oxygen formation energy gets significantly smaller from the 
center to a surface site. In the same manner, the HSE results indicate that 
removal of an oxygen from the surface of a ZnO NW is easier relative to bulk 
cases.

\begin{figure}[htb]
\epsfig{file=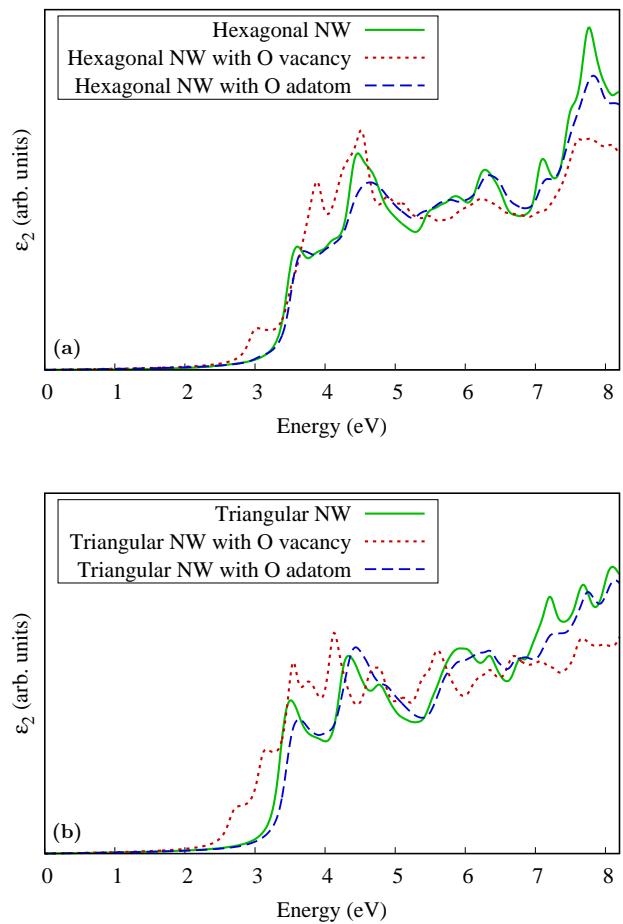,width=8.1cm}
\caption{The optical absorption spectra of (a) the hexagonal and (b) the 
triangular ZnO nanowires calculated using the HSE functional.\label{fig5}}
\end{figure}

The oxygens exposed on the surface show a three-fold coordination with the 
nearest neighbor surface zinc atoms. Removal of an oxygen from the surface of 
the nanowire is possible in the form of a breaking of the bonds 
between the oxygen and its three nearest neighbor zinc atoms. Therefore, an 
oxygen vacancy leaves two excess electrons behind and the charge is 
redistributed in the local region. Because, the Zn-O bonds are broken, these 
excess electrons will have a higher energy relative to their previous bonding 
state when the oxygen was in place. The excess charge is redistributed
around the neighboring Zn and O atoms. A resulting occupied distinct 
state showing partial Zn $3d$ and O $2p$ character due to these excess 
electrons appears in the band gap of ZnO nanowire almost 0.7 eV above the 
valence band as shown in Fig.~\ref{fig4}. These results agree well with the 
experimental findings of Su~\textit{et al.} on the oxygen-deficient ZnO 
nanorods.\cite{Su} The calculated band gap values with the HSE functional are 
2.49 eV for the hexagonal and 2.37 eV for the triangular NWs. The position of 
the gap state in the case of nanowires also agrees well with the previous 
quasiparticle calculations of Lany~\textit{et al.} for the 
V$_{\textrm{\scriptsize O}}^0$ charge-neutral defect state in the bulk 
phase.\cite{Lany2005} This bulk-like behavior of the oxygen vacancy state within 
the band gap of ZnO NW, as being similar for both the hexagonal and triangular 
types, can be related with the stability of the NW structure. In other words, 
oxygen removal on the surface leaves a minimal distortion on the atomic 
positions as seen in Fig.~\ref{fig2}c and Fig.~\ref{fig2}d. The results indicate 
that the wurtzite structure remains to be robust under oxygen deficiencies at 
low concentrations. 

At the HSE level of theory, when the characteristics of the oxygen vacancy 
state compared with respect to the nanowire types, a better localization 
feature is predicted in favor of triangular NW. The DOS plots in Fig.\ref{fig4}a 
and Fig.\ref{fig4}b show the corresponding states with a dispersion over a 
range of 0.5 eV and 0.4 eV for both the hexagonal and triangular cases, 
respectively. Moreover, in the triangular case, existence of the strong peak is 
due to more flat-like nature of the vacancy state.

For the second type of defects considered in this study, an oxygen adatom 
interacts with the nearest neighbor Zn and O atoms and causes
redistribution of the surface charge. Electron transfer from the NWs to the 
adatom allows the formation of new bonds as shown in Fig.~\ref{fig2}a and 
Fig.~\ref{fig2}b for the hexagonal and triangular cases, respectively.
Theory predicts chemisorption with oxygen binding energies of 1.43 eV on the 
hexagonal and 1.46 eV on the triangular NWs. Electronically, the adatom related 
states mainly contribute and shape the edge of the valance band as seen in 
Fig.~\ref{fig4}c and Fig.~\ref{fig4}d. Moreover, presence of an oxygen adatom 
does not lead to a band gap narrowing as opposed to the case of an oxygen 
vacancy. Calculated band gaps are 3.12 eV for the hexagonal and 3.42 eV for the 
triangular NWs. In the latter case, the gap even seems to get a bit larger 
relative to that of the corresponding defect-free NW, which also mimics the 
band gap properties of zinc peroxide (ZnO$_2$) in relation to that of 
ZnO.\cite{Thapa} 

The HSE-calculated optical spectra of the both types of ZnO NWs with oxygen 
vacancies in Fig.~\ref{fig5} show red-shifted absorption features which 
agree well with the experimental observations.\cite{Leiter,Hofmann} In the 
case of the triangular NW, both types of oxygen defects cause a slightly 
larger band gap narrowing relative to their hexagonal NW counterparts. As a 
result, the lowest lying excitation starts at relatively lower energies for the 
triangular NW. Therefore, geometry has an observable effect on the electronic 
and optical structures of the ZnO NWs. 

The oxygen vacancy state contributes largely at the onset of absorption in the 
optical spectra by giving strong vertical excitation features, and therefore, 
is responsible for the reactivity of the NWs in the visible region. In both of 
the NW cases, however, the oxygen adatom slightly blue-shifts the shoulder of 
the absorption spectra as shown in Fig.~\ref{fig5} as being consistent with 
their corresponding electronic band structures. Interestingly, 
Hofmann~\textit{et al.} observed similar characteristics for Zn- and 
O$_2$-annealed ZnO samples.\cite{Hofmann}

\section{CONCLUSION}
Structural geometry optimizations energetically favor the stability of 
free-standing ZnO NWs in the wurtzite structure. The theoretical results show 
that surface oxygen vacancies can occur with formation energies as low as 0.17 
eV for hexagonal and 0.08 eV for triangular-shaped NWs under O-poor conditions. 
On the other hand, an atomic oxygen strongly binds to both types of NWs with 
chemisorption energies of 1.43 eV and 1.46 eV in the hexagonal and triangular 
cases, respectively. In the presence of an isolated O vacancy or adatom, the 
distortion on the atomic positions in the NWs remains local and minimal. 

From a theoretical point of view, the screened exact Coulomb exchange 
contribution in the HSE functional improves the description of electronic and 
optical properties of ZnO NWs over the standard exchange and correlation 
schemes based on the local density approximation. This especially becomes 
important in the presence of oxygen point defects. In fact, a surface oxygen 
vacancy state appears about 0.7 eV above the valance band of both hexagonal and 
triangular NWs. As a consequence, oxygen vacancies significantly red-shift the 
optical absorption threshold of both of the ZnO NWs to the visible region. On 
the other hand, an oxygen adatom forms strong bonds with the ZnO surface giving 
an electronic contribution to the top of the valence band. Resulting optical 
spectra in both of the NW types show a slight blue-shift of the absorption 
shoulder with respect to that of the defect-free NW cases.
Finally, the results indicate that the characteristics of oxygen related point 
defects, electronic and optical properties in particular, show a non-negligible 
dependence on the shape of the NWs.

\end{document}